\documentclass[letterpaper,notoc]{CECS}

\usepackage{graphicx,amsmath,amssymb}
\title{Gravitational Cheshire effect:
       Nonminimally coupled scalar fields may not curve spacetime}

\author{Eloy Ay\'on--Beato${}^{1,2}$, Cristi\'{a}n Mart\'{\i}nez${}^{1}$,
        Ricardo Troncoso${}^{1}$ and Jorge Zanelli${}^{1}$
\footnote{{\it E-mail:} {\tt  ayon-at-cecs.cl, martinez-at-cecs.cl,
                              ratron-at-cecs.cl, jz-at-cecs.cl}}\\
${}^{1}$Centro de Estudios Cient\'{\i}ficos (CECS),
        Casilla 1469, Valdivia, Chile.\\
${}^{2}$Departamento~de~F\'{\i}sica,~CINVESTAV--IPN,
        Apdo. Postal 14--740, 07000, M\'exico D.F., M\'exico.}

\preprint{{\tiny CECS-PHY-05/05}}

\abstract{
It is shown that flat spacetime can be dressed with a real scalar field that
satisfies the nonlinear Klein-Gordon equation without curving spacetime.
Surprisingly, this possibility arises from the nonminimal coupling of the
scalar field with the curvature, since a footprint of the coupling remains in
the energy-momentum tensor even when gravity is switched off.

Requiring the existence of solutions with vanishing energy-momentum tensor
fixes the self-interaction potential as a local function of the scalar field
depending on two coupling constants. The solutions describe shock waves and, in
the Euclidean continuation, instanton configurations in any dimension. As a
consequence of this effect, the tachyonic solutions of the free massive
Klein-Gordon equation become part of the vacuum.}

\begin{document}

\section{Introduction}

According to Einstein's theory of gravitation, gravity is the manifestation of
the curvature of spacetime produced by the presence of matter. In the absence
of sources, the maximally symmetric solution of the Einstein equations
---without cosmological constant--- is Minkowski spacetime. The converse,
however, is not necessarily true: Minkowski space does not imply that spacetime
is devoid of matter. The purpose of this article is to construct a simple and
explicit example where this possibility is realized.

One of the simplest sources is provided by a real scalar field and the
phenomenon described above occurs if the field is nonminimally coupled to
gravity. What is remarkable about this possibility is that if the gravitational
field is switched off, the presence of the nonminimal coupling persists. That
is, if the background geometry is flat, so that the coupling term in the action
vanishes, the footprint of the coupling to gravity is still present in the
energy-momentum tensor. In this sense, the gravitational field plays the role
of Alice's Cheshire cat, where the coupling constant is the grin
\cite{Carroll}.

A scalar field nonminimally coupled to gravity  in $D$ dimensions is described
by the action
\begin{equation}
I=\int {\mathrm{d}^{D}x}\sqrt{-g}\left( \frac{1}{2\kappa }R\;+\frac{1}{2} \phi
\Box \phi -\frac{1}{2}\zeta \,R\,\phi ^{2}-V(\phi )\right) . \label{Action}
\end{equation}
The field equations in flat spacetime reduce to
\begin{equation}
\Box \phi =V'(\phi )\;,  \label{Klein-Gordon}
\end{equation}
and
\begin{equation} \Theta _{\mu \nu }\equiv T_{\mu \nu }+\zeta (\eta _{\mu \nu
}\Box -\nabla _{\mu }\nabla _{\nu })\phi ^{2}=0 \;,  \label{improvement}
\end{equation}
where $T_{\mu \nu}$ is the standard energy-momentum tensor for a minimally
coupled scalar field,
\begin{equation}
T_{\mu \nu }=\partial _{\mu }\phi \partial _{\nu }\phi -\eta_{\mu \nu }\left(
\frac{1}{2}\partial _{\alpha }\phi \partial ^{\alpha }\phi +V(\phi )\right).
\label{Tmn}
\end{equation}

It was observed long ago, that adding an identically conserved
symmetric tensor to (\ref{Tmn}) leads to a theory with improved
renormalizability properties in flat space
\cite{Callan:1970ze,Deser:1970}. The additional term in
Eq.~(\ref{improvement}) is a straightforward generalization of the
``improved" energy-momentum tensor considered there.

For the minimally coupled case $\zeta=0$, Eq.~(\ref{improvement}) reduces to
$T_{\mu \nu}=0$. Any solution of (\ref{Klein-Gordon}) with $T_{\mu \nu}=0$ is
necessarily a constant scalar field $\phi_0$ such that
$V(\phi_0)=V'(\phi_0)=0$. Thus, in this case, the solutions that do not
generate a back reaction are trivial.

In the general case, under the same asymptotic conditions for the scalar field
in flat space as for the conventional tensor (\ref{Tmn}), $\Theta _{\mu \nu }$
gives the same charge generators for the Poincar\'{e} group. However, the
improvement allows for weaker asymptotic conditions, which include a wider
class of solutions with finite charges. In particular, the effect of relaxing
the asymptotic conditions gives rise to the existence of nontrivial solutions
of the nonlinear Klein-Gordon equation with vanishing $\Theta_{\mu \nu}$ for
any nonzero value of $\zeta$.

As it is shown below, requiring the existence of nontrivial solutions of
$\Theta_{\mu \nu}=0$, fixes the self-interaction potential $V(\phi )$ as a
local function of the scalar field depending on two coupling constants, given
by\footnote{This expression is valid for $\zeta \neq 0$ and $\zeta \neq 1/4$.
The case $\zeta =1/4$ is discussed below.}
\begin{equation}
V(\phi)=\frac{2\zeta^2}{(1-4\zeta)^2}\left(\lambda_1
\phi^{\frac{1-2\zeta}{\zeta }} + 8(D-1)(\zeta-\zeta_D)
\lambda_2\phi^{\frac{1}{2\zeta}}\right), \label{Potential}
\end{equation}
where $\zeta_{D}:=\frac{D-2}{4(D-1)}$. Note that if $\zeta =\zeta_{D}$, matter
becomes conformally coupled and the potential has a single coupling constant,
\begin{equation}
V(\phi )=\frac{(D-2)^{2}}{8}\lambda _{1}\phi ^{\frac{2D}{D-2}}\;,
\label{ConfPotential}
\end{equation}
and $\Theta_{\mu \nu}$ becomes traceless, reflecting the scale invariance of
the scalar field equation.

The scalar field configurations satisfying Eq.~(\ref{improvement}) couple to
the gravitational field without curving spacetime. They are found to describe
shock waves and, in the Euclidean continuation, they correspond to instanton
configurations.

\section{Solutions}

Note that Eq.~(\ref{Klein-Gordon}) is a consequence of the conservation of the
energy-momentum tensor provided $\nabla _{\mu }\phi \neq 0$. Therefore, solving
Eq.~(\ref{improvement}) for a nonconstant scalar field is sufficient to satisfy
Eq.~(\ref{Klein-Gordon}). In standard Minkowski coordinates, $x^{\mu}=(t,
x^i)$, $i=1,\ldots ,D-1$, these equations can be written as
\begin{subequations}\label{eq:Tc}
\begin{eqnarray}
\label{eq:Tmunu} 0\!&\!=\!&\!\Theta
_{\mu\nu}=\frac{(2\zeta)^2}{1-4\zeta}\frac{\phi^2}{\sigma}
\partial^2_{\mu\nu}\sigma, \qquad \mu\neq\nu,\\
\label{eq:Tii}
0\!&\!=\!&\!\Theta_{ii}+\Theta_{tt}\!=\!\frac{(2\zeta)^2}{1-4\zeta}\frac{\phi^2}{\sigma}
\left(\partial^2_{ii}\sigma+\partial^2_{tt}\sigma\right) \;\mbox{(no sum)},\\
\label{eq:U}
0\!&\!=\!&\!\Theta_{tt}\!=\!V(\phi)\!-\!\frac{(2\zeta)^2\phi^2}{(1-4\zeta)\sigma}\!
\left[\!\frac{\partial_\alpha\sigma
\partial^\alpha\sigma}{2(1-4\zeta)\sigma}
\!-\!\!\!\sum_{i=1}^{D-1}\partial^2_{ii}\sigma\!\right]\!,\quad\,~
\end{eqnarray}
\end{subequations}
where we have defined
\begin{equation}
\sigma =\phi ^{(4\zeta -1)/2\zeta}\; , \quad (\zeta \neq 0,\;\; \zeta \neq
1/4). \label{eq:sigma}
\end{equation}
It follows from Eq.~(\ref{eq:Tmunu}) that $\sigma $ is completely separable,
\begin{equation}
\sigma =T(t)+X^{1}(x^{1})+\cdots +X^{D-1}(x^{D-1})\;. \label{eq:separ}
\end{equation}
On the other hand, combining Eqs.~(\ref{eq:Tii}), one obtains
\begin{equation}
-\frac{\mathrm{d}^{2}T}{\mathrm{d}t^{2}}=\frac{\mathrm{d}^{2}X^{1}}{\mathrm{d%
}(x^{1})^{2}}=\ldots =\frac{\mathrm{d}^{2}X^{D-1}}{\mathrm{d}(x^{D-1})^{2}}%
=\alpha \;,  \label{eq:alpha}
\end{equation}
where $\alpha $ is a constant. The most general solution of these equations is
\begin{equation}
\sigma =\frac{\alpha }{2}x_{\mu }x^{\mu }+k_{\mu }x^{\mu }+\sigma _{0}\;,
\label{sigma}
\end{equation}
where $k^{\mu }$ and $\sigma _{0}$ are integration constants. The remaining
step is to solve Eq.~(\ref{eq:U}) for $V(\phi )$.

\begin{figure}[h]
\centering
\includegraphics[width=8.5cm]{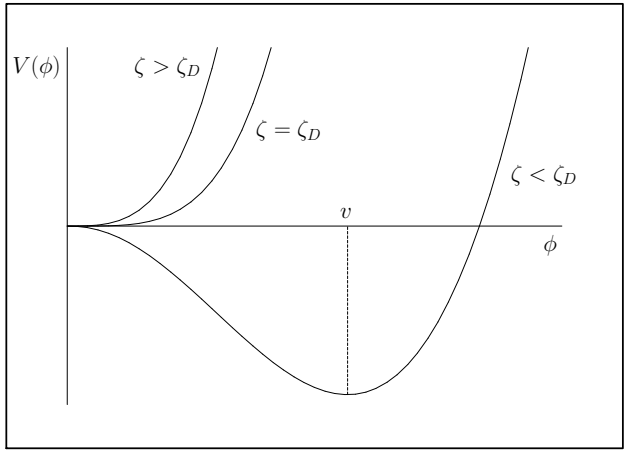}
\caption{\label{fig:pot} For $0<\zeta <1/4$, $\lambda_{1}>0$, and
$\lambda_{2}>0$, the potential has a global minimum at $\phi=0$ if $\zeta \geq
\zeta_{D}$. For $0<\zeta <\zeta_{D}$, it has a local maximum at $\phi=0$ and a
global minimum at $v=\left[ 4(D-1)\lambda_2{\lambda_1}^{-1}(\zeta
-\zeta_{D})/(2\zeta -1)\right] ^{2\zeta /(1-4\zeta)}$. The critical value for
symmetry breaking occurs at the conformal coupling $\zeta =\zeta_D$. For $\zeta
=1/4$ this potential describes again a symmetry breaking self-interaction with
the same qualitative features, where the minimum is now at the vacuum
expectation value $v=\exp [1/2(-\lambda_{1}/\lambda_{2}-D+1)]$.}
\end{figure}

Notably, the solution for the potential turns out to be a local
function of $\phi (x)$ for any nonvanishing value of the parameter
$\zeta \ne 1/4$ given by Eq.~(\ref{Potential}); see also
Fig.~\ref{fig:pot}.

The integration constants $\alpha $, $k^{\mu }$ and $\sigma_{0}$ are not
completely arbitrary, but they are related to the parameters in the potential
by
\begin{eqnarray}
\lambda_{1} &=&k_{\mu }k^{\mu }-2\alpha \sigma _{0}\;,  \nonumber
\label{constants} \\
\lambda_{2} &=&\alpha \;.
\end{eqnarray}
Note that if the solution had been assumed to be invariant under translations
in some direction, $\alpha$ could only be zero. These solutions exist if the
second term in the potential (\ref{Potential}) vanishes, which only occurs for
the special cases $\lambda_2 =0$ or $\zeta=\zeta_D$.

\subsection{Generic cases}

For the class of potentials of the form (\ref{Potential}) with
$\zeta\neq\zeta_D$, the solutions have no nontrivial integration constants and
they are completely determined by the coupling constants of the action,
$\lambda_1$, $\lambda_2$ and $\zeta$. Depending on the value of $\lambda_2$,
the solutions fall into two excluding cases: Lorentz invariant solutions,
including instantons, if $\lambda_{2}\neq 0$, and plane-fronted shock waves for
$\lambda_2=0$.

$\bullet$ $\mathbf{\lambda_{2}\neq 0}$.  After a suitable shift in the
coordinates, the solution reads
\begin{equation}
\phi =\left[ \frac{\lambda _{2}}{2}x^{\mu}x_{\mu} -\frac{\lambda _{1}}{2\lambda
_{2}}\right] ^{-2\zeta /(1-4\zeta )}\;. \label{Phi}
\end{equation}
This solution describes a spherical shock wave, and the scalar field is
singular at $x_{\mu }x^{\mu }=\lambda _{1}/\lambda _{2}^{2}$. In the Euclidean
continuation in spherical coordinates ($ds^{2}=d\rho ^{2}+\rho ^{2}d\Omega
_{D-1}^{2}$), the solution is
\begin{equation}
\widehat{\phi} =\left( \frac{\lambda _{2}}{2}\right) ^{-2\zeta
/(1-4\zeta )}\left[ \rho ^{2}-\frac{\lambda _{1}}{\lambda
_{2}^{2}}\right] ^{-2\zeta /(1-4\zeta )}\;, \label{Instanton}
\end{equation}
so that the singularity at the origin is not present if $\lambda_{1}<0$. This
solution describes an instanton and has finite action for $(D-2)/4D <\zeta
<1/4$,
\begin{eqnarray*}
I_E[\widehat{\phi}] &=&-\int
{\mathrm{d}^{D}x}\sqrt{g}\left(\frac{1}{2} \widehat{\phi}
V^{\prime }(\widehat{\phi}) -V(\widehat{\phi})\right) \;, \\
&=&2^{\frac{2-4\zeta }{1-4\zeta }}\Omega ^{D-1}\;\frac{\zeta
^{2}}{1-4\zeta }\;(-\lambda _{1})^{\frac{D}{2}-\frac{1}{1-4\zeta
}}\;\lambda _{2}^{4\frac{ D-\zeta }{1-4\zeta }(\zeta
-\zeta_D)}\;B\left[ \frac{D}{2},\frac{1}{1-4\zeta
}-\frac{D}{2}\right] \;,
\end{eqnarray*}
where $\Omega ^{D-1}$ is the volume of the sphere $S^{D-1}$, and $B\left[
m,n\right] =\frac{\Gamma (m)\Gamma (n)}{\Gamma (m+n)}$ is the Beta function.

$\bullet \;\mathbf{\lambda _{2}=0}$. In this case one obtains
\begin{equation}
\phi =\left(k_{\mu }x^{\mu }\right) ^{-2\zeta /(1-4\zeta )}\;, \label{Phi'}
\end{equation}
where $k_{\mu }k^{\mu }=\lambda_{1}$. This solution describes a plane-fronted
shock wave traveling along $k_{\mu }$. Note that although the potential
(\ref{Potential}) does not have a mass term proportional to $\phi^{2}$, this
solution has a dispersion relation corresponding to a free particle whose mass
is determined by the self coupling constant, $m^2=-\lambda_1$. Hence the shock
wave is tachyonic for positive
$\lambda_{1}$.\\

\subsection{Special cases}

$\bullet$ $\mathbf{\zeta =\zeta_{D}}$. The conformally coupled case is
particularly interesting. The potential has a single coupling constant and
acquires the form given by Eq.~(\ref{ConfPotential}), which makes the matter
piece of the action (\ref{Action}) conformally invariant. The solution is
regular at infinity since $1/8\leq\zeta_D<1/4$, for $D\geq 3$. Note that in the
generic case discussed above, the solution has no integration constants.
However for $\zeta =\zeta_{D}$, the parameter $\lambda_{2}$ is no longer a
coupling constant because it drops out from the potential. Hence, in the
conformally coupled case, $\lambda_{2}=\alpha $ becomes an integration constant
and the solution (\ref{Phi}) acquires a tunable parameter as expected from
scale invariance. As a consequence, both types of solutions discussed above can
be present now. For $\alpha\neq0$, the solution is given by (\ref{Phi}) with
$\zeta=\zeta_D$,
\begin{equation}\label{ConformalInst}
\phi =\left[ \frac{\alpha }{2}x^{\mu }x_{\mu }-\frac{\lambda
_{1}}{2\alpha } \right] ^{-(D-2)/2}\;,
\end{equation}
which is regular for $\lambda_{1}<0$. The value of the Euclidean action is
finite and given by
\begin{equation}
I_E=(-\lambda_1)^{1-\frac{D}{2}}\frac{(D-2)\sqrt{\pi} \Omega^{D-1} \Gamma
\left(\frac{D}{2}\right)}{4\Gamma\left(\frac{D+1}{2}\right)}\;,
\label{ConfEucAction}
\end{equation}
which is independent from the integration constant $\alpha $. For $\alpha=0$,
the plane-fronted shock wave solution is obtained from (\ref{Phi'}).

$\bullet \; \mathbf{\zeta =1/4}$. In this case, it is useful to
define $\sigma =\ln (\phi )$ and proceed as before. Thus, if
$\lambda_2 \neq 0$, the scalar field is given by
\begin{equation}
\phi =\exp {\left( \frac{\lambda_{2}}{2}x^{\mu}x_{\mu} -\frac{\lambda
_{1}}{2\lambda _{2}}\right) \;}, \label{eq:Psi(x)1/4}
\end{equation}
and the self-interaction potential reads
\begin{equation}
V(\phi )=\frac{\lambda _{2}}{2}\phi ^{2}\left[ 2\ln \phi
+\frac{\lambda_1}{\lambda_2}+D-1\right] .  \label{eq:U(Psi)v1/4}
\end{equation}

As in the generic case, the solution has no nontrivial integration
constants.\\

$\bullet$ $\mathbf{\zeta =1/4}$  \textbf{and} $\mathbf{\lambda_{2}=0}$. In this
case the potential corresponds to a massive term, $V(\phi )=
\frac{1}{2}\lambda_1\phi^{2}$ ($\lambda_1 >0$), and the scalar field is given
by $\phi =\exp (k^{\mu }x_{\mu })$ with $k_{\mu}k^{\mu}=\lambda_1$. This means
that the improvement has the interesting effect of making the tachyonic
solutions of the linear Klein-Gordon equation to have vanishing stress energy
and hence, devoid of any conserved charge.

Note that a real oscillatory solution of the free massive Klein-Gordon equation
in flat spacetime has a non-constant $T_{\mu \nu}$ whereas the improved
$\Theta_{\mu \nu}$ with $\zeta =1/4$ corresponds to dust with constant energy
density.

\section{Summary and discussion}

It has been shown that flat spacetime in $D$ dimensions can be consistently
endowed with nontrivial real scalar fields solving the nonlinear Klein-Gordon
equation without generating back reaction. This is achieved by modifying the
conventional energy-momentum tensor with the addition of a symmetric tensor
that is identically conserved, and which depends on the arbitrary parameter
$\zeta$. This modification can be seen to result from a nonminimal coupling of
the scalar field to gravity: even though this interaction disappears in a flat
background, it generates nontrivial effects through the additional term in the
stress-energy of matter.

It is observed that requiring the existence of nontrivial solutions with
vanishing improved energy-momentum tensor, fixes the self-interaction potential
$V(\phi)$ as a local function of the scalar field, with two arbitrary coupling
constants. The potential exhibits a symmetry breaking transition at the value
of $\zeta=\zeta_D$, for which the matter lagrangian becomes conformally
invariant. In this case, the potential has a single coupling constant and the
solution acquires a tunable parameter.

The space of allowed solutions is enlarged since the improved
energy-momentum tensor permits the use of weaker asymptotic
conditions. This is due to the fact that divergences arising from
the conventional energy-momentum tensor can be canceled by the
contribution of the improvement. This leads to new allowed
configurations having finite charges. In fact, the solutions
presented here have unbounded standard $T_{\mu \nu}$. The stability
of these configurations could be tackled along the lines of
Ref.~\cite{Deser:1983rq}.

The solutions couple to the gravitational field without curving spacetime. They
describe shock waves, and instantons in the Euclidean continuation. For the
free massive Klein-Gordon equation the solutions exist only if $\zeta=1/4$.
These are the tachyons which become part of the spectrum of the (degenerate)
vacuum.

The same results found here would occur in any metric gravitation theory that
admits flat spacetime as a solution. The essential feature is the nonminimal
coupling of the scalar field, while the Einstein-Hilbert term could be replaced
by a more general Lagrangian.

In three dimensions, static and time-dependent configurations on
nontrivial spacetimes without back reaction are known to exist for
scalar fields nonminimally coupled to gravity with negative
cosmological constant
\cite{Natsuume:1999cd,Ayon-Beato:2001sb,Henneaux:2002wm,Gegenberg:2003jr,%
Ayon-Beato:2004ig}.
We have shown here that this effect is neither due to the particular
dimensionality nor to the negative value of the cosmological
constant. It is desirable then to explore whether this phenomenon
may also occur for nontrivial spacetimes in higher dimensions, as
well as in the presence of a non vanishing cosmological constant.

\acknowledgments We thank S. Deser, A. Gomberoff, M. Hassa\"{\i}ne,
M. Loewe, D. Marolf, R. Portugu\'{e}s, and D. Robinson for
enlightening discussions. This work was partially funded by grants
1020629, 1040921, 1051056 and 1051064 from FONDECYT, grants 38495E
and 34222E from CONACyT, and CONICYT/CONACyT Grant 2001-5-02-159.
The generous support of Empresas CMPC to the Centro de Estudios
Cient\'{\i}ficos (CECS) is also acknowledged. CECS is a Millennium
Science Institute and is funded in part by grants from Fundaci\'on
Andes and the Tinker Foundation.

\end{document}